\documentclass[10pt,twocolumn]{ICCAS2024}
 
\usepackage{tabu}
\usepackage{diagbox}
\usepackage{cite}
\usepackage{array}
\usepackage{comment}
\usepackage{anyfontsize}

\begin{document}

\title{CSAC Drift Modeling Considering GPS Signal Quality in the Case of GPS Signal Unavailability}

\author{Seunghyeon Park${}^{1}$ and Joon Hyo Rhee${}^{2*}$ }

\affils{ ${}^{1}$School of Integrated Technology, Yonsei University, \\
Incheon, 21983, Korea (seunghyeon.park@yonsei.ac.kr) \\
${}^{2}$Korea Research Institute of Standards and Science, \\
Daejeon, 34113, Korea (jh.rhee@kriss.re.kr) \\
{\small${}^{*}$ Corresponding author}}


\abstract{
The Global Positioning System (GPS), one of the Global Navigation Satellite Systems (GNSS), provides accurate position, navigation and time (PNT) information to various applications.
One of the application that is highly receiving attention is satellite vehicles, especially Low Earth Orbit (LEO) satellites.
Due to their limited ways to get PNT information and low performance of their onboard clocks, GPS system time (GPST) provided by GPS is a good reference clock to synchronize.
However, GPS is well-known for its vulnerability to intentional or unintentional interference.
This study aims to maintain the onboard clock with less error relative to the GPST even when the GPS signal is disrupted.
In this study, we analyzed two major factors that affects the quality of the GPS measurements: the number of the visible satellites and the geometry of the satellites.
Then, we proposed a weighted model for a Chip-Scale Atomic Clock (CSAC) that mitigates the clock error relative to the GPST while considering the two factors.
Based on this model, a stand-alone CSAC could maintain its error less than 4 $\mu s$, even in a situation where no GPS signals are received for 12 hours.
}


\maketitle


\section{Introduction}

The Global Positioning System (GPS)  \cite{Enge11:Global, DeLorenzo10:WAAS/L5, Chen11:Real, Kim23:Single}, one of the Global Navigation Satellite Systems (GNSS) \cite{Lee23:Seamless, Kim23:Machine, Kim23:Low, Lee24:A, Jia21:Ground}, is being used widely to provide accurate Position, Navigation and Timing (PNT) service for various applications. 
One of the application that is highly receiving attention is PNT on the satellite vehicles, especially Low Earth Orbit (LEO) satellites.
Due to their limited ways to get PNT information, GPS is an easy and simple solution for them.
In general, onboard clocks on the satellites are an Oven Controlled Crystal Oscillators (OCXO) or a compact atomic clocks such as Chip-Scale Atomic Clocks (CSAC) because the satellites can usually afford a limited weight and volume to load \cite{Coffer99:Long}.
The performance of the onboard clocks (e.g. stability, accuracy, or drift rate) is relatively lower than that of the atomic clocks on the ground \cite{Wang15:Precise}. 
To maintain the onboard clock stable and accurate, the satellite should be able to compare the time difference between its clock and a reference clock \cite{Enge99:Local, DeLorenzo10:WAAS/L5, Kim24:Development, Lee22:Performance_Estimation}.
By the comparison, the satellite could synchronize its clock or compensate the error of the clock to maintain its accuracy and stability as needed \cite{Farina10:A}.
The reference clock must be more accurate and more stable than the onboard clock, and GPS system time (GPST) is an easy option to use as an accurate, stable, and reliable reference clock. 
Several studies \cite{Wang15:Precise, Yang21:GNSS, Arceo09:Optimal, Zuohu10:High, Lee23:Phase, Lee23:Improved, Lee21:Performance} have been conducted to ensure clock synchronization when GPS system time is consistently available.

However, GPS is also well-known as its vulnerability to intentional signal attacks such as jamming and spoofing due to its low power of received signals.
Additionally, GPS signals can be disrupted by high-power radio frequency interference (RFI) \cite{Schmidt20:A, Park21:Single, Park18:Dual, Kim19:Mitigation, Park17:Adaptive, Jeong20:RSS} and ionospheric anomalies  \cite{Jiao15:Comparison, Sun20:Performance,  Seo11:Availability, Lee22:Optimal, Sun21:Markov, Lee17:Monitoring, Ahmed17:Statistical}.
To keep the satellite's clock accurate and stable even when the GPS is not available, one could model the time error of its onboard clock refer to the GPST without valid GPS signals.
If the time error could be estimated precisely based on the model, the satellite user can adjust or steer its onboard clock to keep it accurate and stable compared to the case of free running \cite{Hanson71:Clock}.

In this paper, we analyzed the effect on the received GPST that can be caused by two factors: number of the visible satellites and geometry of the satellites.
The simulation environment, process, and the result of the analysis is introduced in Section 2.
In Section 3, we suggested a clock drift model which is not only reflecting the clock's own characteristics but also considering the two factors.
To verify the performance of the suggested model, we performed an experiment with simulated GPS signals and a CSAC. 
The result is given in Section 4.
Section 5 concludes this paper.

\section{Analysis of the GPST obtained by a GPS receiver}

A user can get PNT information from the GPS by receiving the GPS signals from multiple GPS satellites over the sky.
Position $[x,y,z]$ and time $[t]$ are resolved from the pseudoranges, which are calculated by measuring these signals.
The measurements contain errors which could be originated from various error sources and factors.
In this study, we focused on two factors that significantly affect the measurement errors.

One of the two factors is the number of the visible satellites.
Theoretically, a user would be able to get PNT information when four signals are received in minimum.
However, there are various scenarios where some signals, though received, may not be usable.
A signal from a GPS satellite can be disrupted by intentional or unintentional noise, which lowers its SNR at the receiver.
In this case, even though the signal is received and acquired by the receiver, the pseudorange measurement could be too noisy to use, thereby affecting the quality of the final PNT solution.
For users operating at high altitudes, such as LEO satellites, the number of visible satellites may be lower than that for ground users.
Thus, the issue of having fewer visible satellites arises more frequently for these users \cite{Gong14:The}. 

Another factor that affects the error is the geometry of the satellites.
Geometry refers to the directional distribution of the satellites relative to the receiver's location.
It can be represented numerically by the Dilution of Precision (DOP) factor \cite{Tahsin15:Analysis}. 
To summarize, we assumed that the positioning and timing errors are primarily affected by the number of visible satellites and their geometry \cite{Joardar16:Analyses}.
Therefore, we initially created an environment to examine how the quality of GPS measurements varies with these two factors.
The result of this experiment is used to suggest a weighted model that reflects the quality of GPS measurements in Section 3.

\subsection{Experiment setting}

We used Spectracom's GNSS simulator (GSG-63) to generate scenarios with the desired number of satellites and their geometry.
The simulator allows us to select the date, location, and the maximum number of satellites as needed.
We created three independent scenarios, and the conditions for each scenario are shown in Table \ref{tab:conditions}.

\begin{table}
\centering
\caption{Initial conditions of three independent scenarios.}
\label{tab:conditions}
\scriptsize
\begin{tabular}{|l|c|c|c|} 
\hline
Scenario &   (1)   &   (2)   &   (3) \\\hline
Starting Date&Jan. 10, 2023 & Jan. 8, 2023 & Jan. 9, 2023 \\\hline
Starting Time&10:00&04:00&04:00 \\\hline
Duration&6h&6h&6h \\\hline
Latitude & 37.6315$^\circ$ & 43.0830$^\circ$ & 43.7800$^\circ$ \\\hline
Longitude & 126.3633$^\circ$ & -77.5890$^\circ$ & 11.2500$^\circ$ \\\hline
Altitude & 11.665 m & 206.550 m & 31.200 m \\\hline
Max no. of satellites & 7 & 8 & 9 \\\hline
\end{tabular}
\end{table}

\begin{figure} 
    \centering 
    \includegraphics[width=1.0\linewidth]{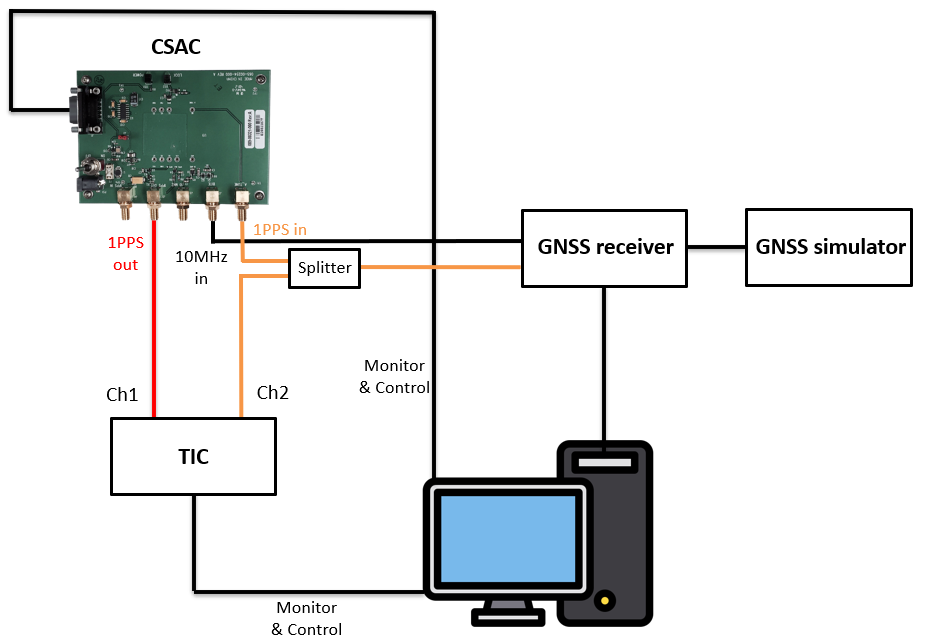}
    \caption{Overall configuration of the hardware devices of our experiment.} 
    \label{fig:configuration}
\end{figure}

For example, the first scenario is created with starting date and time of 10:00:00 AM on January 10, 2023, latitude of 37.6315$^\circ$, longitude of 126.3633$^\circ$, altitude of 11.665 m, and the maximum number of satellites is limited to 7.
The GNSS simulator generates a simulated signal based on the given scenario, and the simulated signal is connected to the input port of the GNSS receiver (Novatel's PwrPak7).
The GNSS receiver produces a position and time solution from the simulated signal and it gives out a 1PPS signal that is synchronized with the time information.
On the other hand, we used Microchip's SA65 as a stand-alone free-running CSAC.
The CSAC has an input port that can receive an external 1PPS signal to synchronize with.
To synchronize the CSAC with the external 1PPS signal, we used an external PC connected to the CSAC which can monitor and control the CSAC.
So the PC can send a command to perform the synchronization of the CSAC.
The synchronization with GPS receiver's 1PPS output signal was only performed at the beginning of each experiment.
The CSAC also has an output port that can provide its 1PPS signal to another device.
We used a Time Interval Counter (TIC, Keysight's 53220A) to measure the time difference between the 1PPS output signal from the GNSS receiver and the 1PPS output signal from the CSAC.
The overall configuration of the hardware devices of our experiment is shown in Fig. \ref{fig:configuration}.

\begin{figure} 
    \centering 
    \includegraphics[width=1.0\linewidth]{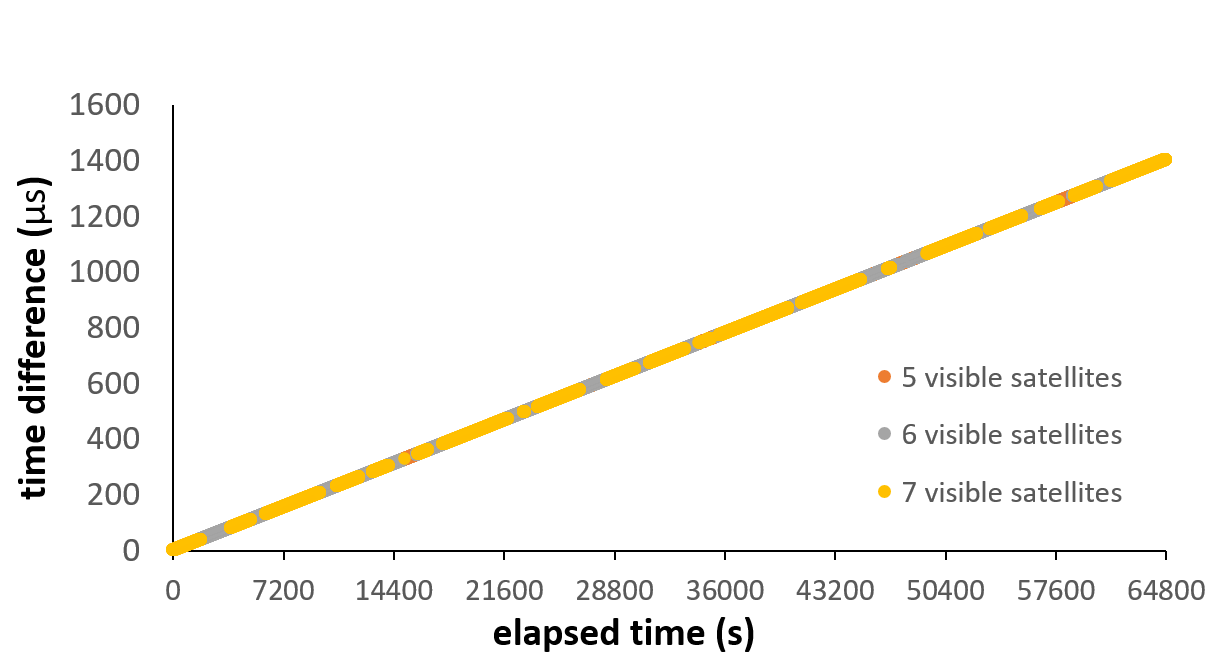}
    \caption{Time difference measurements between the 1PPS output signal from the CSAC and the 1PPS output signal from the GNSS receiver.} 
    \label{fig:ticresult}
\end{figure}

For each scenario, we produced 5 datasets under the same conditions to ensure consistency.
An example case of the experiment is plotted in Fig. \ref{fig:ticresult}.
This shows the time difference measurements between the 1PPS output of the GNSS receiver and the 1PPS output of the CSAC after time synchronization.
Each experiment was performed for 18 hours.

\subsection{Quality change due to the number of visible satellites}

The quality metric ``noise variance ($\sigma^2$)'' \cite{Zhang19:Novel, Kutty11:Noise} was adopted to quantify the variation in the quality of measured values based on the number of visible GPS satellites.
This metric checks whether the same scenario produces consistent time drift values when reproduced multiple times.
For a given set of clock drift measurements $X = (x_1,x_2,...,x_n)$, the noise variance is defined as follows \cite{Zhang12:Some}:
\begin{equation}
    \sigma^2 = \frac{1}{ \binom n2} \sum_{i=1}^{n-1} \sum_{j=i+1}^n (x_i-x_j)^2
\end{equation}
This equation calculates the variance by considering every pair of datasets.
This approach involves several datasets for the same simulator-generated scenario, ensuring that reliable and consistent measurements are produced for each scenario.
Variations can occur in the measurement and comparison of this noise value, even under identical conditions.
A low noise variance is considered as an indicator of good measurement quality because it represents the variability between multiple datasets obtained under the same conditions.

Clock drift measurements of 6-hour duration, initiated with a synchronization command, were used for this experiment.
The synchronization command was applied after configuring scenario (1) from Table \ref{tab:conditions} with the GNSS simulator.
The same process was repeated 5 times to generate 5 measurement datasets for scenario (1).
The difference, or what we call noise, between each pair of datasets was computed for every 10 combinations among the 5 datasets.
Each dataset was divided by the number of visible satellites, allowing us to calculate the variance of the noise based on the number of visible satellites.

The results shown in Table \ref{tab:numsats} indicate that as the number of visible satellites increases, the dispersion of noise between datasets decreases.
Therefore, it can be concluded that the accuracy of timing measurements improves with greater satellite visibility.

\begin{table}
\centering
\caption{Quality of measurements based on the number of visible satellites.}
\label{tab:numsats}
\begin{tabular}{|l|c|c|c|} 
\hline
visible satellites &   5   &   6   &   7 \\\hline
noise variance ($\sigma^2$) & 0.7164 & 0.5504 & 0.4812 \\\hline
\end{tabular}
\end{table}

\begin{figure} 
    \centering 
    \includegraphics[width=0.8\linewidth]{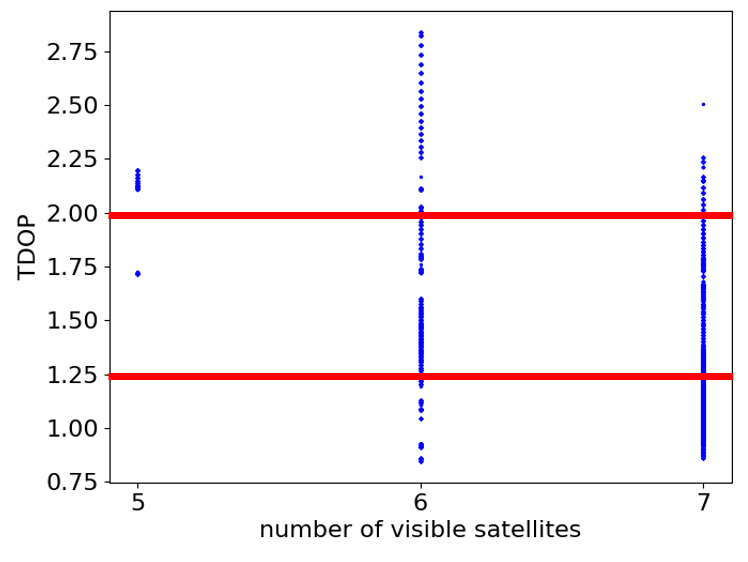}
    \caption{Relationship between visible number of satellites and TDOP.} 
    \label{fig:relationship}
\end{figure}

\subsection{Quality change due to dilution of precision (DOP)}

The Dilution of Precision (DOP) reflects the relationship between pseudorange error and user location/timing error.
The DOP decreases when the positional allocation of visible satellites is more uniformly spread.
Given the focus of this study on timing precision, Time Dilution of Precision (TDOP), which relates to the user's timing error, was primarily used.
To examine the distribution of TDOP, the association between the number of visible satellites and TDOP in the initial measurement dataset is displayed in Fig. \ref{fig:relationship}.
This showed that TDOP values ranged roughly between 0.8 and 3.0. 
Therefore, we divided TDOP into three ranges with certain thresholds for intuitive analysis: less than 1.25, between 1.25 and less than 2.0, and 2.0 or greater. 
The noise between two datasets was computed by evaluating every combination from the five datasets using the same procedure as previously described, to examine quality changes in relation to TDOP.
The results, categorized into three ranges---TDOP less than 1.25, TDOP between 1.25 and less than 2.0, and TDOP of 2.0 or greater---are shown in Fig. \ref{fig:relationship}. 

It was observed that the noise variance between datasets under the same conditions decreases with a lower TDOP.
Consequently, it can be concluded that the quality of timing error measurement improves with a lower TDOP.

\begin{table}
\centering
\caption{Quality of measurements based on TDOP values.}
\label{tab:tdop}
\begin{tabular}{|l|c|c|c|} 
\hline
&    \small TDOP$<$1.25   & \small 1.25$\leq$TDOP$<$2   & \small  2$\leq$TDOP \\\hline
$\sigma^2$ & 0.451973 & 0.563479 & 0.722449 \\\hline
\end{tabular}
\end{table}

\section{CSAC drift modeling considering GPS signal quality}

\subsection{Model selection}
The Chip-Scale Atomic Clock (CSAC) demonstrates a good short-term stability, especially given its low power consumption ($<$120 mW) and compact size ($<$16 cc) \cite{DeNatale08:Compact}.
Nonetheless, to enhance its long-term accuracy and stability, timing synchronization to GPS System Time (GPST) is necessary.
Effective timing synchronization can be achieved by using a clock drift model that describes the drift characteristics of the CSAC relative to GPST.

Regression models are among the most effective and straightforward techniques for predicting a target variable on a continuous scale \cite{Specht91:general, Li17:Adaptive, Yang03:Adaptive, Specht94:Experience}.
Therefore, identifying the clock drift pattern through regression is crucial, as it addresses one of the key challenges in machine learning involving continuous values.

\begin{figure} 
    \centering 
    \includegraphics[width=1.0\linewidth]{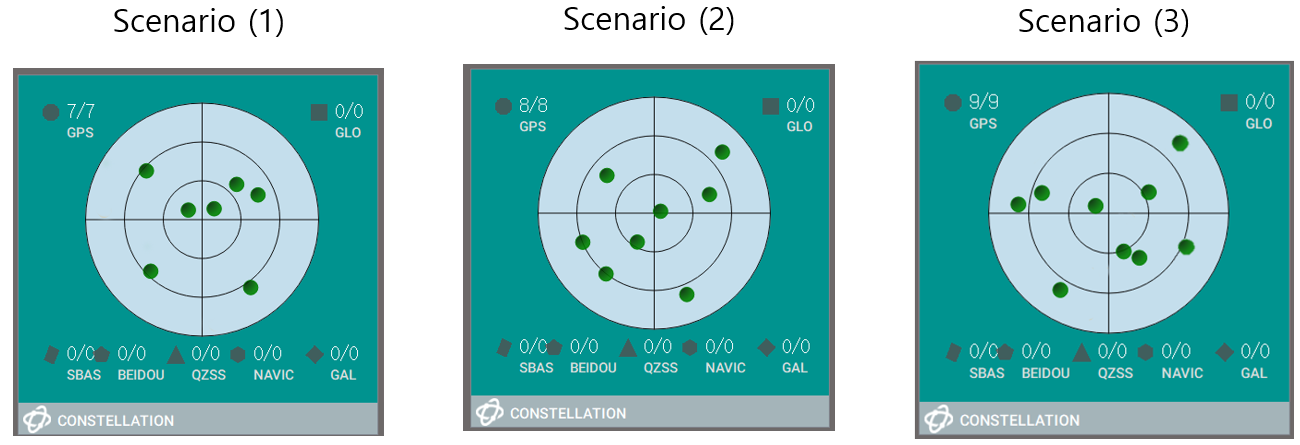}
    \caption{Snapshot of initial GPS simulator satellite constellation for 3 scenarios (Incheon, New York, Florence).} 
    \label{fig:scenario}
\end{figure}

Previous research \cite{Zhan16:Utilization} suggested that the timing drift of CSAC for GPST can be modeled linearly.
To determine whether the CSAC for GPS signals produced by simulators follows the same pattern, an experiment was designed for this study.
Moreover, maintaining clock synchronization even when GPS signals are subsequently lost is critical.
Thus, we evaluated the coasting ability of the model, rather than just its fit with initial measurements.
We utilized a model fitted to clock drift values measured every 2 seconds for the first 6 hours to forecast clock drift for the subsequent 12 hours.
This model was applied in three distinct, arbitrarily created scenarios, as depicted in Fig. \ref{fig:scenario}.

In particular, the coasting performance of various clock drift models---such as linear, and polynomial (second, third, and fourth degree) models---fitted to the initial observed values was compared to determine the most appropriate fundamental model.
Table \ref{tab:without} displays the outcomes of coasting with each model by calculating the root mean square error (RMSE) between the predicted and actual values.

This comparison revealed that the linear equation model produced the lowest RMSE, at 866.68 ns.
Consequently, the study focused on refining this model by incorporating weights derived from the first-order model.
It was previously established that GPS signal quality declines as the TDOP value increases or the number of visible satellites decreases.
Thus, `the reciprocal of TDOP' and `the ratio of the number of visible satellites to the maximum number of visible satellites' were used to weight the linear model in this study.

\subsection{Experiment setting}

For the purpose of analyzing coasting performance, three scenarios were created using a GPS simulator.
Upon applying the newly suggested technique, performance enhancements were examined for each scenario.
The initial satellite setups for measuring each scenario are depicted in Fig. \ref{fig:scenario}.
The scenarios vary by date, location, and maximum number of visible satellites.
Next, a time synchronization command was applied via the PC to synchronize CSAC time with GPST.
Even at the point of synchronization, there was a time difference of approximately 4 $\mu s$ due to the atomic clock's performance and the cable distance between the terminals.
Since GPST synchronization was performed only once during the experiment, the clock error of the CSAC continued to increase over time, influenced by the characteristics of the CSAC.

\begin{table}
\centering
\caption{Coasting performance of model without considering quality.}
\label{tab:without}
\begin{tabular}{|l|c|c|c|} 
\hline
Scenario &   (1)   &   (2)   &   (3) \\\hline
RMSE & 3.825 $\mu s$ & 0.866 $\mu s$ & 2.136 $\mu s$ \\\hline
\end{tabular}
\end{table}

\begin{figure} 
    \centering 
    \includegraphics[width=1.0\linewidth]{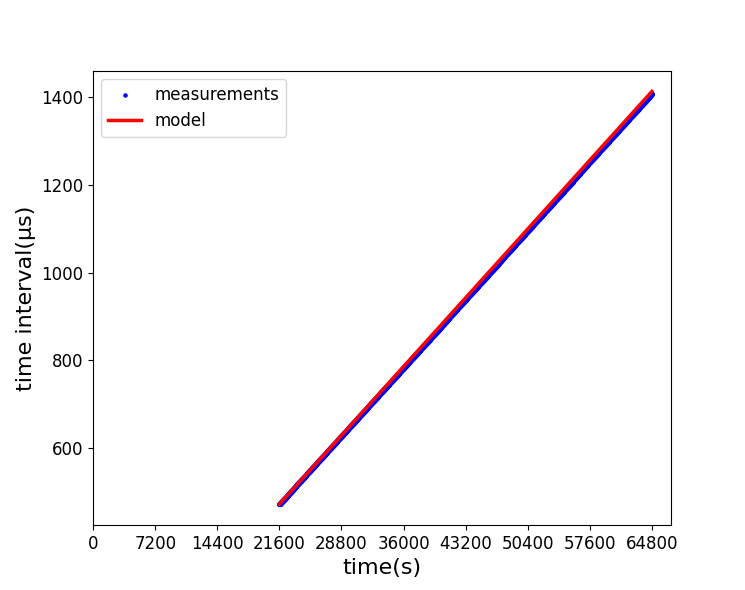}
    \caption{Coasting of weighted (using TDOP) linear model of t\_CSAC minus GPST.} 
    \label{fig:tdopcoasting}
\end{figure}

After GPST synchronization, the 1PPS output signal from the CSAC was connected to channel 1 of the Time Interval Counter (TIC), and the time difference between CSAC and GPS was measured by connecting the 1PPS signal from GPST to channel 2.
The time difference at the TIC was calculated for each pulse by subtracting the pulse rise time of the GPS signal from the pulse rise time of the CSAC 1PPS signal.
For 18 hours straight, this time difference was monitored every 2 seconds, yielding a total of 32,401 data points for each scenario.
Out of these, the model was built using only the first 10,801 measurements from the 6-hour period.
Fig. \ref{fig:tdopcoasting} exhibits the model weighted by the reciprocal of TDOP for scenario (1).

\section{Results}

To achieve clock coasting with only the Chip-Scale Atomic Clock (CSAC), the rising edge time of the CSAC's 1PPS signal must be subtracted from the clock drift predicted by the model.
Consequently, the root mean square error (RMSE) between the actual and the predicted clock drift values was used to examine the coasting performance.
The data collected during the first six hours was used to build the model.
Table \ref{tab:final} displays the time coasting performance results using this model alone over the subsequent 12 hours.
The findings indicate that in all three scenarios, the RMSE value decreased when the weighted model took quality factors into account.
Moreover, performance improved significantly when weighted by `the reciprocal of TDOP' rather than by `the ratio of the number of visible satellites to the maximum number of visible satellites.'
Notably, the experiments indicate that even if GPS signals are not received for 12 hours, CSAC alone can accurately predict GPS time with errors ranging from 0.7 to 4 $\mu s$.

\begin{table}
\centering
\caption{Quality of measurements based on TDOP values.}
\label{tab:final}
\resizebox{\columnwidth}{!}{\begin{tabular}{|p{1.5cm}|c|c|c|} 
\hline
\diagbox[dir=SE,width=1.92cm,height=1.3cm]{Scenario}{Model} &   Linear   &   Weighted (satnum)    &   Weighted (TDOP) \\\hline
(1) & 3.825 $\mu s$ & 3.774 $\mu s$ & 3.548 $\mu s$ \\\hline
(2) & 0.866 $\mu s$ & 0.827 $\mu s$ & 0.775 $\mu s$ \\\hline
(3) & 2.136 $\mu s$ & 2.002 $\mu s$ & 1.930 $\mu s$ \\\hline
\end{tabular}}
\end{table}

\section{Conclusion}

This study proposes a weighted model that mitigates clock error relative to GPST, even when the GPS signal is disrupted.
Moreover, this model considers not only the clock's intrinsic characteristics but also the quality of GPS measurements.
Interestingly, the results show that a stand-alone CSAC could reliably maintain its error at less than 4 $\mu s$ in our experiments, even when no GPS signals were received for 12 hours.

\section*{ACKNOWLEDGEMENT}

This work was supported by the Institute of Information \& Communications Technology Planning \& Evaluation (IITP) grant funded by the Korean government (MSIT) (RS-2022-00143911, AI Excellence Global Innovative Leader Education Program).
This research was also supported in part by the Future Space Navigation and Satellite Research Center through the National Research Foundation of Korea (NRF) funded by the Ministry of Science and ICT (MSIT), Republic of Korea under Grant 2022M1A3C2074404, and in part by the NRF funded by the Korean government (MSIT) under Grant RS-2024-00358298.

\bibliographystyle{IEEEtran}
\bibliography{mybibfile, IUS_publications}

\end{document}